\documentclass[aps, pre, showpacs, superscriptaddress, reprint]{revtex4-1}
\usepackage{graphicx}
\usepackage{color}
\usepackage{amssymb}
\usepackage{amsmath}
\usepackage{tabularx}
\usepackage{bm}
\usepackage{ltxtable}
\usepackage{natbib} 
\usepackage{longtable}
\usepackage{float}

\newcommand{\bB}{{\bf B}}

\newcommand{\be}{{\bf e}}
\newcommand{\bE}{{\bf E}}

\newcommand{\bj}{\textbf{j}}
\newcommand{\bq}{\textbf{q}}
\newcommand{\bu}{\textbf{u}}

\newcommand{\ii}{\mathrm{i}}

\begin{document}


\title[]{Axisymmetric dynamo action is possible with anisotropic conductivity}

\author{Franck Plunian}
\email{Franck.Plunian@univ-grenoble-alpes.fr}
\affiliation{Universit\'e Grenoble Alpes, Universit\'e Savoie Mont Blanc, CNRS, IRD, IFSTTAR, ISTerre, 38000 Grenoble, France}

\author{Thierry Alboussi\`ere}
\email{Thierry.Alboussiere@ens-lyon.fr}
\affiliation{Univ. Lyon, UCBL, ENS de Lyon, CNRS, UMR 5276 LGL-TPE, F-69622 LYON, France}
 
\date{\today}

\begin{abstract}
A milestone of dynamo theory is Cowling's theorem, known in its modern form as the impossibility for an axisymmetric velocity field to generate an axisymmetric magnetic field  by dynamo action. Using an anisotropic electrical conductivity we show that an axisymmetric dynamo is in fact possible with a motion as simple as solid body rotation. On top of that the instability analysis can be conducted entirely analytically, leading to an explicit expression of the dynamo threshold which is the only example in dynamo theory.  
\end{abstract}

\maketitle

\section{Introduction}
Since the pioneering study of \citet{Cowling1934} there has been a constant effort to improve the demonstration of the so-called Cowling's (antidynamo) theorem. In its modern form this theorem states that an axisymmetric magnetic field cannot be generated by dynamo action under the assumption of axisymmetry of velocity field, electrical conductivity, magnetic permeability and shape of the conductor \cite{Moffatt1978,Ivers1984,Fearn1988,Proctor2007,Kaiser2014}. Cowling's theorem encompasses time-dependent flows \cite{Backus1957,Braginskii1964}, non-solenoidal flows and variable conductivity \cite{Hide1982,Lortz1982}. However nothing has yet been said about the effect of an anisotropic electrical conductivity and how in this case Cowling's theorem is overcome. 
A demonstration of dynamo action with shear and anisotropic conductivity has already been given \cite{Ruderman1984}, but for a different geometry and within asymptotic limits relevant to the fast dynamo problem.

Beyond its theoretical interest, this issue is relevant to at least three fields of physics. In astrophysics it is well known that, in the mean-field approximation, an anisotropic tensor of magnetic diffusivity may naturally occur from anisotropic gradients of magnetohydrodynamic turbulence \cite{Krause1980}. 
In plasma physics, just like thermal conductivity \cite{Onofri2010},  the electrical conductivity in the magnetic field direction is different from the electrical conductivity in the direction perpendicular to the magnetic field \cite{Braginskii1965}. This usually occurs in a plasma which is already magnetized. Although this does not preclude dynamo action we will not examine this issue here, considering that there is no external magnetic field. 
Finally, as will be shown below, a dynamo experiment can be designed on the basis of our anisotropic conductivity model. The results show that such an experiment is feasible, which is welcome because experimental dynamo demonstrations are rather rare. 

\section{Anisotropic conductivity}
Let us consider a material of electrical conductivity $\sigma$ such that $\sigma=\sigma_1$ in a given direction $\bq$, and $\sigma=\sigma_0\ge \sigma_1$  in the directions perpendicular to $\bq$. 
We choose $\bq$ as a unit vector in the horizontal plane,
 \begin{equation}
 \bq=\cos\alpha \; \be_r+\sin\alpha \; \be_{\theta},
 \end{equation}
 where $(\be_r,\be_{\theta},\be_z)$ is a cylindrical coordinate system and $\alpha$ a constant angle. 
 In a companion paper another choice for $\bq$, within a cartesian frame, is studied \cite{Alboussiere2019}.
 
\begin{figure}
\includegraphics[scale=0.29]{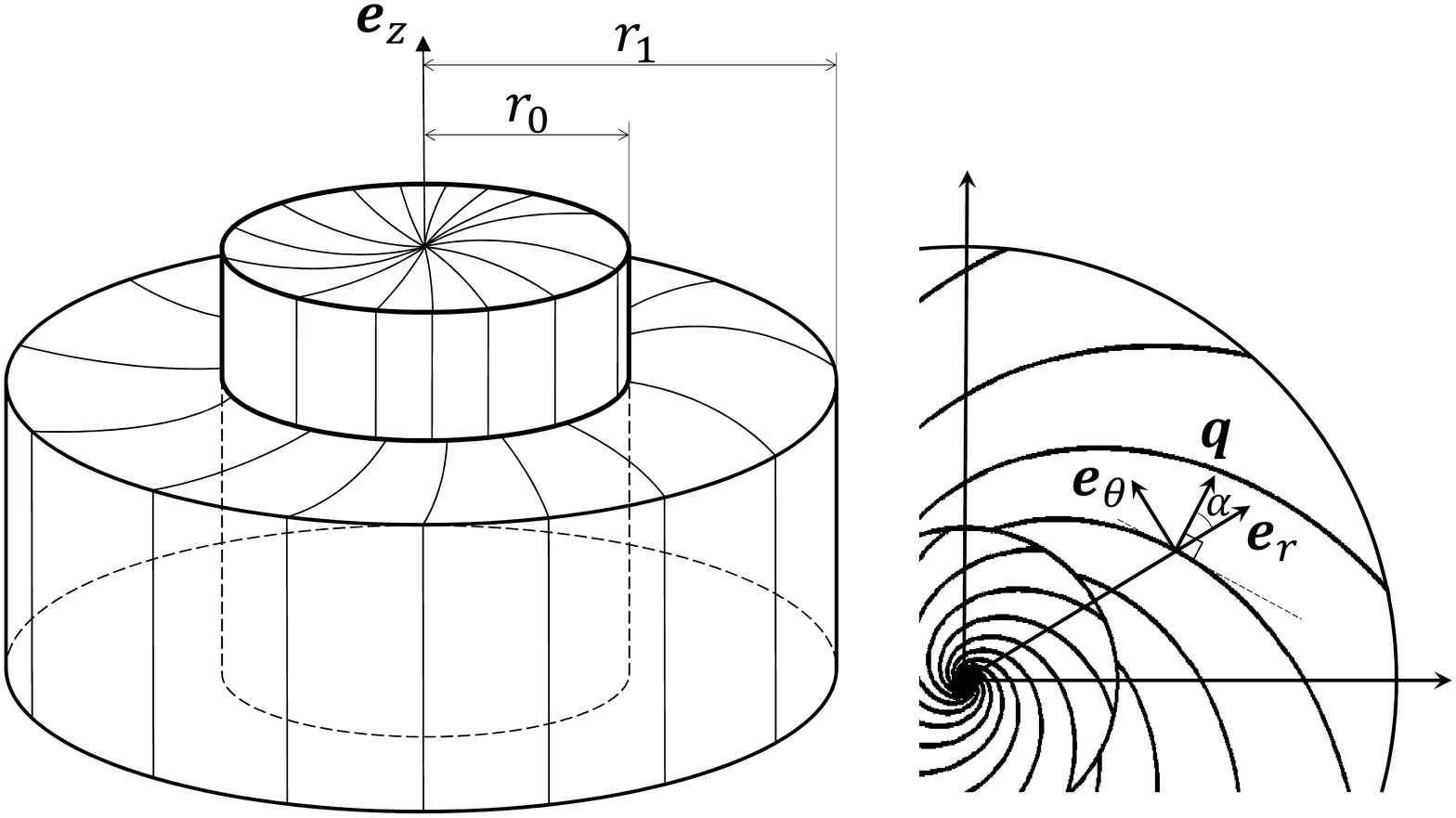}
\caption{Left: The inner-cylinder of radius $r_0$ rotates as a solid-body within an outer-cylinder at rest. The radial boundary $r_1$ of the outer-cylinder is rejected at infinity. In the limit $\eta_1=\infty$
the electric currents follow logarithmic spiral trajectories. 
Right: The vector $\bq$, which makes a constant angle $\alpha$ with the radial direction, is perpendicular to the spiraling current.}
\label{fig:spirales}
\end{figure}

 In Fig.~\ref{fig:spirales} the curved lines correspond to the directions of the large conductivity $\sigma_0$. They are perpendicular to $\bq$ and describe logaritmic spirals.

Writing Ohm's law $\bj=\sigma_1 \bE$ in the direction of $\bq$, and $\bj=\sigma_0 \bE$ in the directions perpendicular to $\bq$, leads to the following conductivity tensor 
\begin{equation}
\sigma_{ij}=\sigma_0 \delta_{ij} + (\sigma_1-\sigma_0)q_i q_j.
\label{eq:conductivity tensor}
\end{equation}
Inversing (\ref{eq:conductivity tensor}) leads to the resistivity tensor \cite{Ruderman1984}
\begin{equation}
R_{ij}=\frac{1}{\sigma_0} \delta_{ij} + (\frac{1}{\sigma_1}-\frac{1}{\sigma_0})q_i q_j.
\label{eq:resistivity tensor}
\end{equation}
We consider the solid body rotation $\bu$
of a cylinder of radius $r_0$ embedded in an infinite medium at rest (Fig.~\ref{fig:spirales}), both regions having the same resistivity tensor $R_{ij}$. 

The magnetic induction $\bB$ satisfies the equation
\begin{equation}
\partial_t \bB = \nabla \times (\bu \times \bB) - \nabla \times \left( [\eta] \nabla \times \bB \right),
\label{eq:induction equation}
\end{equation}
where $[\eta]$ is the magnetic diffusivity tensor defined as $\eta_{ij}=R_{ij}/\mu_0$, $\mu_0$ being the magnetic permeability of vacuum.
Renormalizing the distance, magnetic diffusivity and time by respectively $r_0,(\mu_0\sigma_0)^{-1}$ and $\mu_0\sigma_0r_0^2$, the dimensionless form of the induction equation is identical to (\ref{eq:induction equation}), but
with 
\begin{equation}
\eta_{ij}=\delta_{ij}+\eta_1 q_iq_j, \;\;\; \eta_1=\frac{\sigma_0}{\sigma_1}-1,
\label{eq:dimensionless diffusivity}
\end{equation}
and
\begin{equation}
\bu=
\left\{
\begin{split}
r\Omega \be_{\theta}&,&r< 1 \\
0&,&r>1
\end{split}\;\;\;,
\right. \label{eq:velocity}
\end{equation}
where $\Omega$ is the dimensionless angular velocity of the inner-cylinder.

\section{Resolution}
Provided the velocity is stationary and $z$-independent, an axisymmetric magnetic induction can be searched in the form
\begin{equation}
\bB(r,z,t)= \tilde{B}\be_{\theta} + \nabla \times \left( \tilde{A}\be_{\theta}\right),
\end{equation}
with $(\tilde{A}, \tilde{B})=(A, B)\exp(\gamma t + \ii kz)$ where $\gamma$ is the instability growthrate, $k$ the vertical wavenumber of the corresponding eigenmode, and
where $A$ and $B$ depend only on the radial coordinate $r$. 
Thus the magnetic induction takes the form
\begin{equation}
\bB=\left(-\ii k A, B, \frac{1}{r}\partial_r(rA) \right)\exp(\gamma t + \ii kz),
\label{eq:magnetic induction}
\end{equation}
dynamo action corresponding to $\Re\{\gamma\}>0$.

From (\ref{eq:velocity}) and (\ref{eq:magnetic induction})  we find that $\nabla\times(\bu\times\bB)=0$ in each region $r<1$ and $r>1$.
Replacing (\ref{eq:dimensionless diffusivity}) and (\ref{eq:velocity}) in the induction equation (\ref{eq:induction equation}) leads to
\begin{eqnarray}
\gamma A + D_k(A) &=& \ii \eta_1 cs k B - \eta_1s^2 D_k(A) \label{eq:gamma A}\\
\gamma B + D_k(B) &=& - \ii \eta_1csk D_k(A) - \eta_1c^2 k^2 B   \label{eq:gamma B},
\end{eqnarray}
where $D_{\nu}(X)=\nu^2X-\partial_r\left(\frac{1}{r}\partial_r(rX)\right)$, $c=\cos\alpha$ and $s=\sin\alpha$.  
 
Looking for stationary solutions the dynamo threshold corresponds to $\gamma=0$. 
Then the system (\ref{eq:gamma A}-\ref{eq:gamma B}) implies
\begin{equation}
D_{\tilde{k}}(B) = D_k(B-\ii\frac{ck}{s}A)=0,
\end{equation}
where 
\begin{equation}
\tilde{k}=k\left(\frac{1+\eta_1}{1+\eta_1s^2}\right)^{1/2}.
\end{equation}

The solutions of $D_{\nu}(X)=0$ being a linear combination of $I_1(\nu r)$ and $K_1(\nu r)$, we find
\begin{eqnarray}
r< 1,&&
\left\{
\begin{split}
&A= \frac{s}{\ii ck}\left( \lambda\frac{I_1(\tilde{k}r)}{I_1(\tilde{k})} + \mu \frac{I_1(kr)}{I_1(k)}\right) \\
&B=\lambda\frac{I_1(\tilde{k}r)}{I_1(\tilde{k})}
\end{split}
\right. \label{eq:ABrle1}\\
r> 1,&&
\left\{
\begin{split}
&A= \frac{s}{\ii ck}\left( \lambda\frac{K_1(\tilde{k}r)}{K_1(\tilde{k})} + \mu \frac{K_1(kr)}{K_1(k)}\right)\\
&B=\lambda\frac{K_1(\tilde{k}r)}{K_1(\tilde{k})},
\end{split}
\right. \label{eq:ABrge1}
\end{eqnarray}
where $I_1$ and $K_1$ are modified Bessel functions of first and second kind. In (\ref{eq:ABrle1}) and (\ref{eq:ABrge1}) the following boundary conditions have been applied to $A$ and $B$:
finite values at $r=0$, continuity at $r=1$, and $\lim\limits_{r \to \infty}A,B=0$.

From (\ref{eq:magnetic induction}), the continuity of $\bB$ is satisfied provided
 $\partial_rA$ is also continuous at $r=1$. From (\ref{eq:ABrle1}) and (\ref{eq:ABrge1}) this leads to the following identity between $\lambda$ and $\mu$
 \begin{equation}
 \lambda \Gamma(\tilde{k}) + \mu \Gamma(k)=0,
 \label{eq:A'cont}
 \end{equation}
 with
 \begin{equation}
 \Gamma(x)=x\left(\frac{I_0(x)}{I_1(x)} + \frac{K_0(x)}{K_1(x)}\right) \equiv \left(I_1(x) K_1(x)\right)^{-1},
 \label{eq:Gamma}
 \end{equation}
the last equality coming from the Wronskian relation $I_m(x)K_{m+1}(x)+I_{m+1}(x)K_m(x)=1/x$.

\begin{figure}[H]
\includegraphics[scale=0.3]{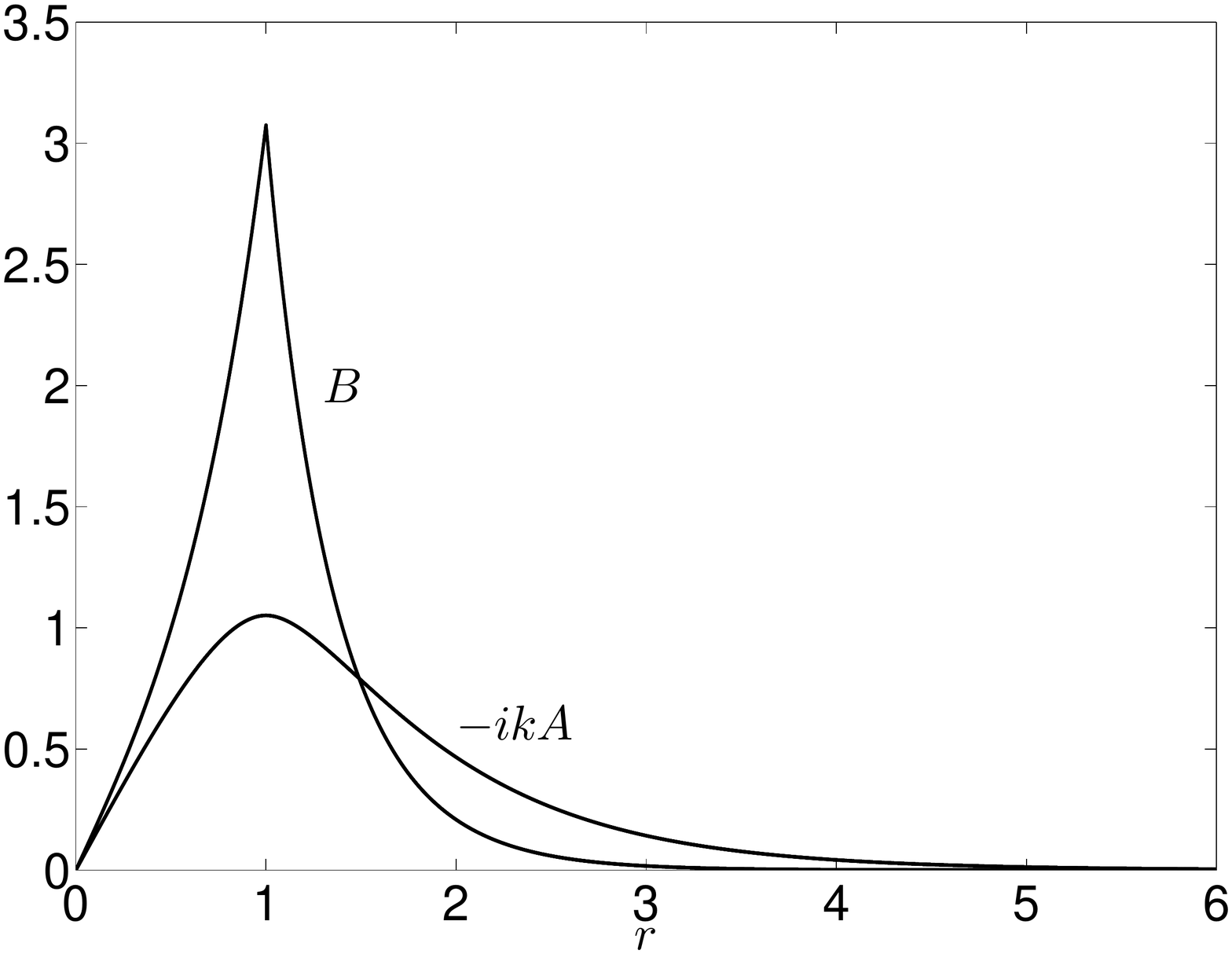}
\caption{Eigenmodes $-\ii kA$ and $B$ versus $r$ for $\eta_1=\infty, k=1.1, \alpha=0.16\pi$, $\lambda=\Gamma(k)$ and $\mu=-\Gamma(\tilde{k})$.}
\label{fig:eigenmodes}
\end{figure}

In Fig. \ref{fig:eigenmodes} the eigenmodes $-\ii kA$ and $B$ are plotted versus $r$ for $\lambda=\Gamma(k)$ and $\mu=-\Gamma(\tilde{k})$ such that (\ref{eq:A'cont}) is satisfied. Both $-\ii kA$ and $B$ reach their maximum at $r=1$.

\begin{figure}[h]
\includegraphics[scale=0.32]{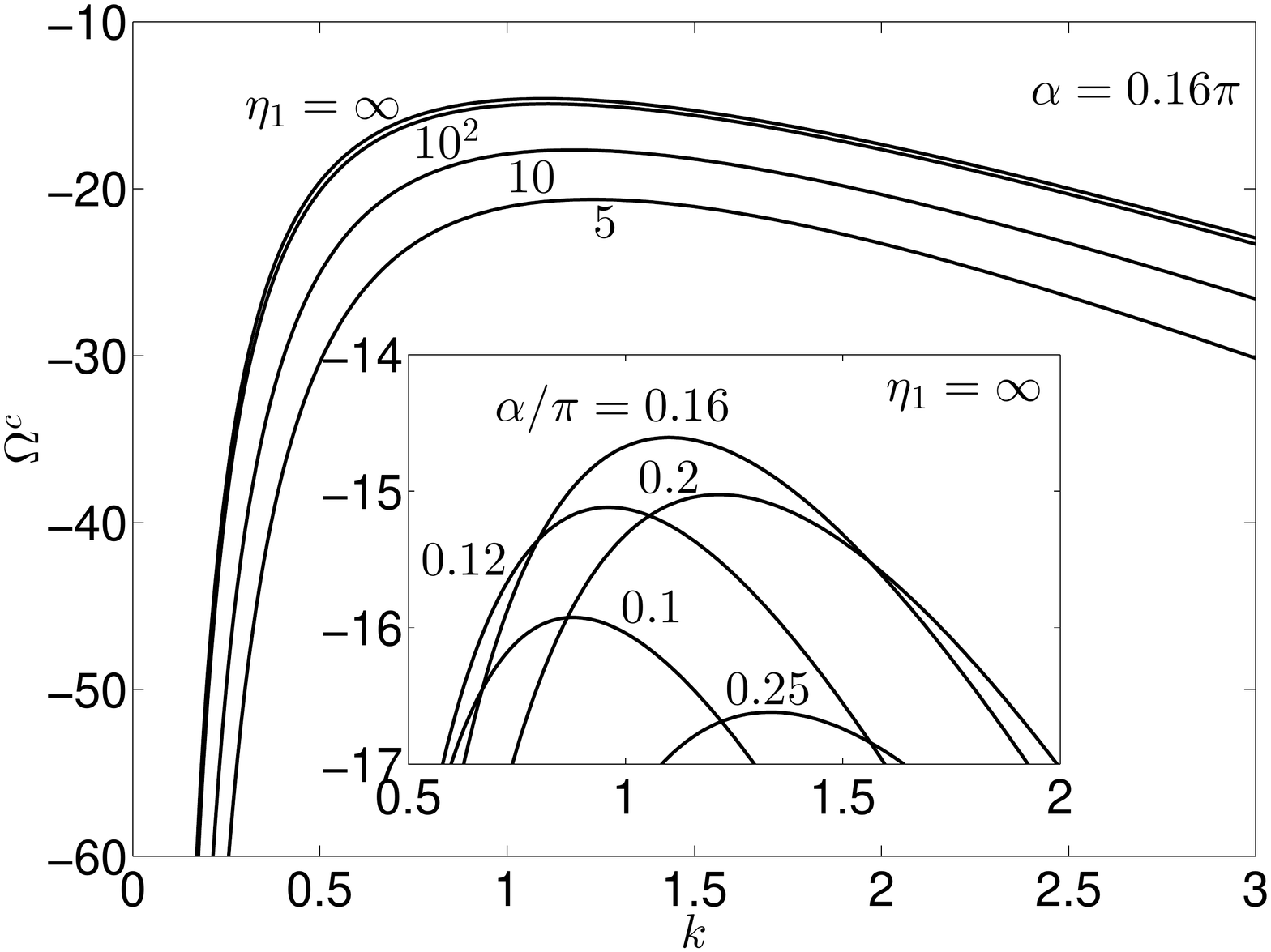}
\caption{Curves of the dynamo threshold, $\Omega^c$ versus $k$, for $\alpha=0.16\pi$ and $\eta_1= 5, 10, 10^2, \infty$. Inset: $\eta_1=\infty$ and $\alpha/\pi=0.1, 0.12, 0.16, 0.2, 0.25$.}
\label{fig:marginal curves}
\end{figure}

Finally, the 
tangential components of the electric field $\bE=- \bu \times \bB + [\eta]\nabla\times \bB $ have to be continuous at $r=1$. The continuity of  $\bE_z$ implies the following identity
\begin{equation}
(\partial_r B -\ii k \Omega A)(r=1^-) = \partial_r B(r=1^+).
\label{eq:Ezcont}
\end{equation}
According to Fig. \ref{fig:eigenmodes}, from which we have $-\ii kA\ge0$, $\partial_r B(r<1)>0$ and $\partial_r B(r>1)\le0$, the only way to satisfy (\ref{eq:Ezcont}) is to have $\Omega<0$.
Replacing (\ref{eq:ABrle1}), (\ref{eq:ABrge1}) and (\ref{eq:A'cont}) in (\ref{eq:Ezcont}) leads to the dynamo threshold
\begin{equation}
\Omega^c = \frac{c}{s}\left(I_1(\tilde{k}) K_1(\tilde{k})-I_1(k)K_1(k)\right)^{-1}.
 \end{equation}
As previously noted we find negative values of $\Omega^c$, dynamo action corresponding to $|\Omega|\ge |\Omega^c|$.  In Fig. \ref{fig:marginal curves} the curves of the dynamo threshold are plotted for different values of $\eta_1$ and $\alpha$. The minimum value of $|\Omega^c|$ is obtained for $\eta_1 \rightarrow \infty$, $k^*=1.1$ and  $\alpha^*=0.16 \pi$,
 \begin{equation}
 \Omega^*=\min_{\eta,k,\alpha}|\Omega^c|=14.61.
 \end{equation}

\section{Dynamo mechanism}
\begin{figure}
\includegraphics[scale=0.3,angle=-90]{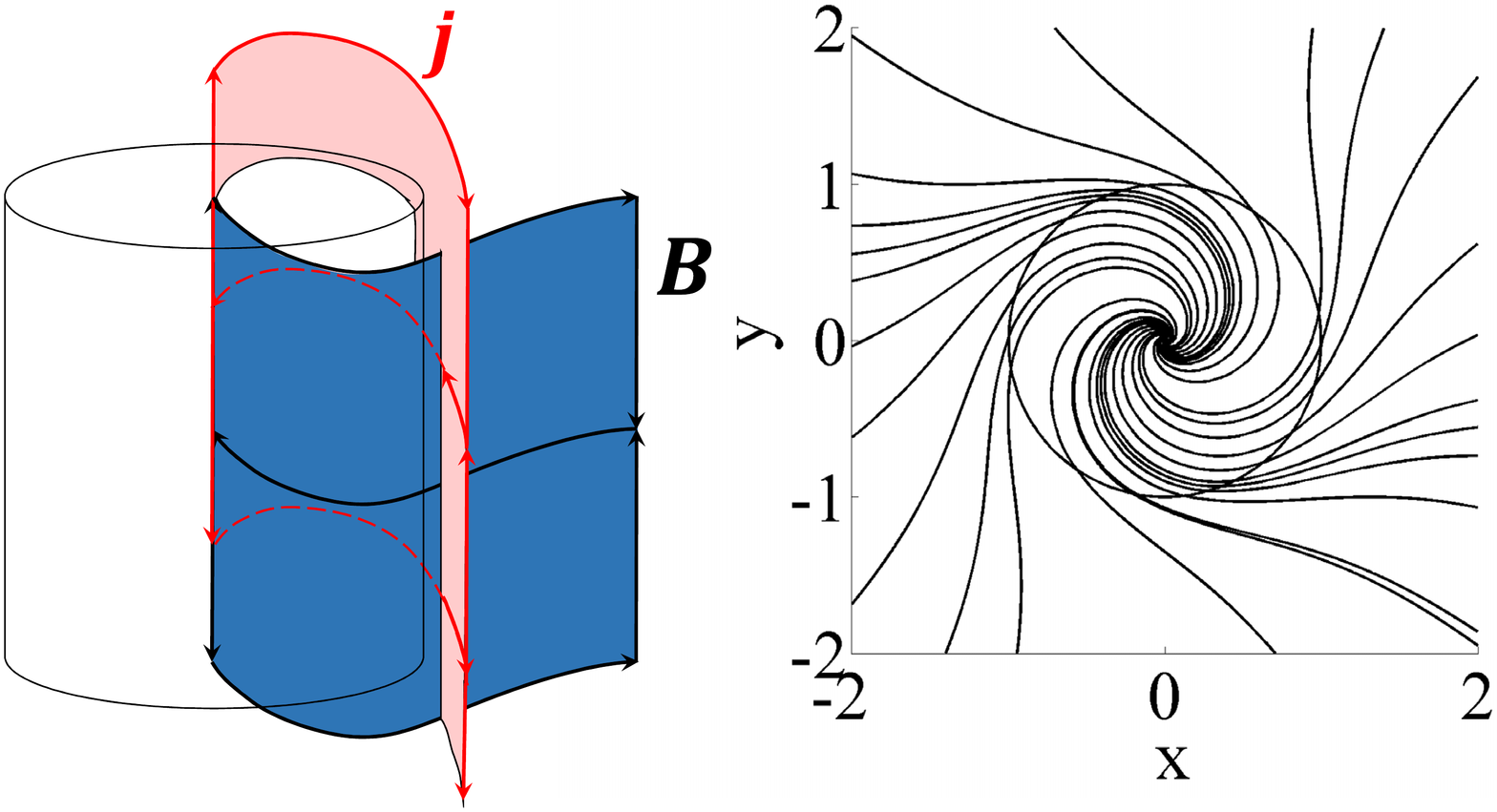}
\caption{Left: Three dimensional sketch of some trajectories of the current density $\bj$ and the magnetic field $\bB$. 
Right: Magnetic field lines in the horizonthal plane $z=0$. The magnetic field is distorded by the differential rotation while the current density is bent by the conductivity anisotropy.}
\label{fig:dynamo mechanism}
\end{figure}

The dynamo mechanism can be 
described  as a two step process as illustrated in Fig. \ref{fig:dynamo mechanism}. The boundary condition (\ref{eq:Ezcont}) implies that $\bB_{\theta}$ is generated from $\bB_r$ by differential rotation between the inner and outer cylinders. This leads to distorsion of magnetic field lines as shown in the right of Fig.~\ref{fig:dynamo mechanism}.
In return the first term on the right hand side of (\ref{eq:gamma A}) corresponds to the generation of $\bB_r$ from $\bB_{\theta}$, provided $\eta_1csk\ne0$. This appears more clearly rewritting (\ref{eq:gamma A}-\ref{eq:gamma B}) as
\begin{eqnarray}
\gamma \bB_r &=& \eta_1 cs k^2 \bB_{\theta} - (1+\eta_1s^2) D_k(\bB_r) \label{eq:Br}\\
\gamma \bB_{\theta}  &=& \eta_1cs D_k(\bB_r) - (D_k+\eta_1c^2 k^2) \bB_{\theta}   \label{eq:Bteta}.
\end{eqnarray}
In the left of Fig.~\ref{fig:dynamo mechanism} the horizonthal currents are represented to follow the direction of logaritmic spirals. 
To show it, the current density $\bj=\nabla \times \bB$ is written in the form
\begin{equation}
\bj=\left(-\ii k B, D_k(A), \frac{1}{r}\partial_r(rB) \right)\exp(\gamma t + \ii kz).
\label{eq:current density}
\end{equation}
From (\ref{eq:gamma A}) taken at the threshold $\gamma=0$, we find that
\begin{equation}
j_{\theta}=-\frac{\eta_1cs}{1+\eta_1s^2} j_r,
\label{eq:current spirals}
\end{equation}
corresponding to the equation of logaritmic spirals. In the limit $\eta_1\rightarrow \infty$ we find that $\bj\cdot\bq=0$, the currents following the trajectories given 
in Fig.\ref{fig:spirales}.

Dynamo action thus occurs through differential rotation conjugated to anisotropic diffusion. 
 For $\eta_1=0$ (isotropic diffusion)  or $cs=0$, in (\ref{eq:Br}-\ref{eq:Bteta}) $\bB_r$ and $\bB_{\theta}$ are decoupled, canceling any hope of dynamo action in accordance with Cowling's theorem.
 
 It is interesting to note that in (\ref{eq:Br}) and (\ref{eq:Bteta}), in each equation it is the first term on the right-hand side which helps for dynamo action. These terms correspond to the off-diagonal coefficients of the anisotropic diffusivity tensor (\ref{eq:dimensionless diffusivity}). 
Therefore the diagonal and off-diagonal coefficients act respectively against and in favour of dynamo action. 

%
%

\section{Conclusions}
The neutral point argument of Cowling relies on the impossibility, in an axisymmetric configuration, of maintaining a toroidal current density
\cite{Cowling1934}. 
This argument falls as soon as the conductivity is a tensor because, in this case, the cross product of a toroidal velocity field with a poloidal magnetic field can actually produce a toroidal current density. In other words, the anistropic conductivity forces the current density to follow spiraling trajectories, with nonzero azimuthal components, thus overcoming Cowling's theorem. 

Beyond the fact that with an anisotropic conductivity an axisymmetric dynamo can be operated from a simple solid-body rotation, it is interesting to put some numbers on the previous results.
Considering an inner-cylinder of radius $r_0=0.05$m, taking the conductivity of copper $\mu_0 \sigma_0 \approx 72.9$s.m$^{-2}$, leads to a dynamo threshold $f^*=\Omega^*(2\pi\mu_0\sigma_0r_0^2)^{-1} \approx 12.8$Hz. Provided the cylinder height and outer radius $r_1$ are sufficiently large, this is experimentally achievable. 
Such an anisotropic conductivity can be easily manufactured by alternating thin layers of two materials with different conductivities and a logarithmic spiral arrangement of these thin layers. 
Of course, the resulting conductivity is no longer homogeneous and, more importantly, it does not satisfy the axisymetry hypothesis of Cowling's theorem. 
However, provided the layers are thin enough, an anisotropic conductivity model is relevant to design such a dynamo experiment.   
Another dynamo experiment design with spiraling wires has been studied \cite{Priede2013}. Though the geometry is different, the dynamo threshold is comparable to the present one.

\bibliographystyle{apsrev4-1}

\begin{thebibliography}{16}%
\makeatletter
\providecommand \@ifxundefined [1]{%
 \@ifx{#1\undefined}
}%
\providecommand \@ifnum [1]{%
 \ifnum #1\expandafter \@firstoftwo
 \else \expandafter \@secondoftwo
 \fi
}%
\providecommand \@ifx [1]{%
 \ifx #1\expandafter \@firstoftwo
 \else \expandafter \@secondoftwo
 \fi
}%
\providecommand \natexlab [1]{#1}%
\providecommand \enquote  [1]{``#1''}%
\providecommand \bibnamefont  [1]{#1}%
\providecommand \bibfnamefont [1]{#1}%
\providecommand \citenamefont [1]{#1}%
\providecommand \href@noop [0]{\@secondoftwo}%
\providecommand \href [0]{\begingroup \@sanitize@url \@href}%
\providecommand \@href[1]{\@@startlink{#1}\@@href}%
\providecommand \@@href[1]{\endgroup#1\@@endlink}%
\providecommand \@sanitize@url [0]{\catcode `\\12\catcode `\$12\catcode
  `\&12\catcode `\#12\catcode `\^12\catcode `\_12\catcode `\%12\relax}%
\providecommand \@@startlink[1]{}%
\providecommand \@@endlink[0]{}%
\providecommand \url  [0]{\begingroup\@sanitize@url \@url }%
\providecommand \@url [1]{\endgroup\@href {#1}{\urlprefix }}%
\providecommand \urlprefix  [0]{URL }%
\providecommand \Eprint [0]{\href }%
\providecommand \doibase [0]{http://dx.doi.org/}%
\providecommand \selectlanguage [0]{\@gobble}%
\providecommand \bibinfo  [0]{\@secondoftwo}%
\providecommand \bibfield  [0]{\@secondoftwo}%
\providecommand \translation [1]{[#1]}%
\providecommand \BibitemOpen [0]{}%
\providecommand \bibitemStop [0]{}%
\providecommand \bibitemNoStop [0]{.\EOS\space}%
\providecommand \EOS [0]{\spacefactor3000\relax}%
\providecommand \BibitemShut  [1]{\csname bibitem#1\endcsname}%
\let\auto@bib@innerbib\@empty
\bibitem [{\citenamefont {Cowling}(1934)}]{Cowling1934}%
  \BibitemOpen
  \bibfield  {author} {\bibinfo {author} {\bibfnamefont {T.~G.}\ \bibnamefont
  {Cowling}},\ }\href@noop {} {\bibfield  {journal} {\bibinfo  {journal} {Mon.
  Not. R. Astr. Soc.}\ }\textbf {\bibinfo {volume} {94}},\ \bibinfo {pages}
  {39} (\bibinfo {year} {1934})}\BibitemShut {NoStop}%
\bibitem [{\citenamefont {Moffatt}(1978)}]{Moffatt1978}%
  \BibitemOpen
  \bibfield  {author} {\bibinfo {author} {\bibfnamefont {H.~K.}\ \bibnamefont
  {Moffatt}},\ }\href@noop {} {\emph {\bibinfo {title} {Magnetic Field
  Generation in Electrically Conducting Fluids}}},\ edited by\ \bibinfo
  {editor} {\bibfnamefont {E.}~\bibnamefont {Cambridge University~Press},
  \bibfnamefont {Cambridge}}\ (\bibinfo {year} {1978})\BibitemShut {NoStop}%
\bibitem [{\citenamefont {Ivers}\ and\ \citenamefont
  {James}(1984)}]{Ivers1984}%
  \BibitemOpen
  \bibfield  {author} {\bibinfo {author} {\bibfnamefont {D.~J.}\ \bibnamefont
  {Ivers}}\ and\ \bibinfo {author} {\bibfnamefont {R.~W.}\ \bibnamefont
  {James}},\ }\href@noop {} {\bibfield  {journal} {\bibinfo  {journal} {Philos.
  Trans. Roy. Soc. London Ser. A}\ }\textbf {\bibinfo {volume} {312}},\
  \bibinfo {pages} {179–218} (\bibinfo {year} {1984})}\BibitemShut {NoStop}%
\bibitem [{\citenamefont {Fearn}\ \emph {et~al.}(1988)\citenamefont {Fearn},
  \citenamefont {Roberts},\ and\ \citenamefont {Soward}}]{Fearn1988}%
  \BibitemOpen
  \bibfield  {author} {\bibinfo {author} {\bibfnamefont {D.}~\bibnamefont
  {Fearn}}, \bibinfo {author} {\bibfnamefont {P.}~\bibnamefont {Roberts}}, \
  and\ \bibinfo {author} {\bibfnamefont {A.}~\bibnamefont {Soward}},\ }in\
  \href@noop {} {\emph {\bibinfo {booktitle} {Pitman Research Notes in
  Mathematics Series 168}}},\ \bibinfo {editor} {edited by\ \bibinfo {editor}
  {\bibfnamefont {G.}~\bibnamefont {Galdi}}\ and\ \bibinfo {editor}
  {\bibfnamefont {B.}~\bibnamefont {Straughan}}}\ (\bibinfo  {publisher}
  {Longman Scientific and Technical},\ \bibinfo {address} {New York, USA},\
  \bibinfo {year} {1988})\ pp.\ \bibinfo {pages} {60--324}\BibitemShut
  {NoStop}%
\bibitem [{\citenamefont {Proctor}(2007)}]{Proctor2007}%
  \BibitemOpen
  \bibfield  {author} {\bibinfo {author} {\bibfnamefont {M.}~\bibnamefont
  {Proctor}},\ }in\ \href@noop {} {\emph {\bibinfo {booktitle} {Mathematical
  Aspects of Natural Dynamos}}},\ \bibinfo {editor} {edited by\ \bibinfo
  {editor} {\bibfnamefont {E.}~\bibnamefont {Dormy}}\ and\ \bibinfo {editor}
  {\bibfnamefont {A.}~\bibnamefont {Soward}}}\ (\bibinfo  {publisher} {Chapman
  and Hall/CRC},\ \bibinfo {address} {Boca Raton, USA},\ \bibinfo {year}
  {2007})\ pp.\ \bibinfo {pages} {18--41}\BibitemShut {NoStop}%
\bibitem [{\citenamefont {Kaiser}\ and\ \citenamefont
  {Tilgner}(2014)}]{Kaiser2014}%
  \BibitemOpen
  \bibfield  {author} {\bibinfo {author} {\bibfnamefont {R.}~\bibnamefont
  {Kaiser}}\ and\ \bibinfo {author} {\bibfnamefont {A.}~\bibnamefont
  {Tilgner}},\ }\href@noop {} {\bibfield  {journal} {\bibinfo  {journal} {SIAM
  Journal on Applied Mathematics}\ }\textbf {\bibinfo {volume} {74}},\ \bibinfo
  {pages} {571} (\bibinfo {year} {2014})}\BibitemShut {NoStop}%
\bibitem [{\citenamefont {Backus}(1957)}]{Backus1957}%
  \BibitemOpen
  \bibfield  {author} {\bibinfo {author} {\bibfnamefont {G.}~\bibnamefont
  {Backus}},\ }\href@noop {} {\bibfield  {journal} {\bibinfo  {journal}
  {Astrophys. J.}\ }\textbf {\bibinfo {volume} {125}},\ \bibinfo {pages}
  {500–524} (\bibinfo {year} {1957})}\BibitemShut {NoStop}%
\bibitem [{\citenamefont {Braginskii}(1964)}]{Braginskii1964}%
  \BibitemOpen
  \bibfield  {author} {\bibinfo {author} {\bibfnamefont {S.}~\bibnamefont
  {Braginskii}},\ }\href@noop {} {\bibfield  {journal} {\bibinfo  {journal}
  {Soviet Phys. JETP}\ }\textbf {\bibinfo {volume} {20}},\ \bibinfo {pages}
  {1462} (\bibinfo {year} {1965})}\BibitemShut {NoStop}%
\bibitem [{\citenamefont {{Hide}}\ and\ \citenamefont
  {{Palmer}}(1982)}]{Hide1982}%
  \BibitemOpen
  \bibfield  {author} {\bibinfo {author} {\bibfnamefont {R.}~\bibnamefont
  {{Hide}}}\ and\ \bibinfo {author} {\bibfnamefont {T.~N.}\ \bibnamefont
  {{Palmer}}},\ }\href {\doibase 10.1080/03091928208208961} {\bibfield
  {journal} {\bibinfo  {journal} {Geophysical and Astrophysical Fluid
  Dynamics}\ }\textbf {\bibinfo {volume} {19}},\ \bibinfo {pages} {301}
  (\bibinfo {year} {1982})}\BibitemShut {NoStop}%
\bibitem [{\citenamefont {Lortz}\ \emph {et~al.}(1982)\citenamefont {Lortz},
  \citenamefont {Meyer-Spasche},\ and\ \citenamefont {T\"ornig}}]{Lortz1982}%
  \BibitemOpen
  \bibfield  {author} {\bibinfo {author} {\bibfnamefont {D.}~\bibnamefont
  {Lortz}}, \bibinfo {author} {\bibfnamefont {R.}~\bibnamefont
  {Meyer-Spasche}}, \ and\ \bibinfo {author} {\bibfnamefont {W.}~\bibnamefont
  {T\"ornig}},\ }\href {\doibase 10.1002/mma.1670040107} {\bibfield  {journal}
  {\bibinfo  {journal} {Mathematical Methods in the Applied Sciences}\ }\textbf
  {\bibinfo {volume} {4}},\ \bibinfo {pages} {91} (\bibinfo {year}
  {1982})}\BibitemShut {NoStop}%
\bibitem [{\citenamefont {Ruderman}\ and\ \citenamefont
  {Ruzmaikin}(1984)}]{Ruderman1984}%
  \BibitemOpen
  \bibfield  {author} {\bibinfo {author} {\bibfnamefont {M.~S.}\ \bibnamefont
  {Ruderman}}\ and\ \bibinfo {author} {\bibfnamefont {A.~A.}\ \bibnamefont
  {Ruzmaikin}},\ }\href {\doibase 10.1080/03091928408210135} {\bibfield
  {journal} {\bibinfo  {journal} {Geophysical \& Astrophysical Fluid Dynamics}\
  }\textbf {\bibinfo {volume} {28}},\ \bibinfo {pages} {77} (\bibinfo {year}
  {1984})}\BibitemShut {NoStop}%
\bibitem [{\citenamefont {{Krause}}\ and\ \citenamefont
  {R\"adler}(1980)}]{Krause1980}%
  \BibitemOpen
  \bibfield  {author} {\bibinfo {author} {\bibfnamefont {F.}~\bibnamefont
  {{Krause}}}\ and\ \bibinfo {author} {\bibfnamefont {K.~H.}\ \bibnamefont
  {R\"adler}},\ }\ \href@noop {} {\emph {\bibinfo {booktitle} {Mean-Field Magnetohydrodynamics and Dynamo Theory}}}, (\bibinfo  {publisher} {Pergamon
  Press},\ \bibinfo {address} {Oxford},\ \bibinfo {year}
  {1980})\BibitemShut {NoStop}%
\bibitem [{\citenamefont {Onofri}\ \emph {et~al.}(2010)\citenamefont {Onofri},
  \citenamefont {Malara},\ and\ \citenamefont {Veltri}}]{Onofri2010}%
  \BibitemOpen
  \bibfield  {author} {\bibinfo {author} {\bibfnamefont {M.}~\bibnamefont
  {Onofri}}, \bibinfo {author} {\bibfnamefont {F.}~\bibnamefont {Malara}}, \
  and\ \bibinfo {author} {\bibfnamefont {P.}~\bibnamefont {Veltri}},\
  }\href@noop {} {\bibfield  {journal} {\bibinfo  {journal} {Phys. Rev. Lett.}\
  }\textbf {\bibinfo {volume} {105}},\ \bibinfo {pages} {215006} (\bibinfo
  {year} {2010})}\BibitemShut {NoStop}%
\bibitem [{\citenamefont {{Braginskii}}(1965)}]{Braginskii1965}%
  \BibitemOpen
  \bibfield  {author} {\bibinfo {author} {\bibfnamefont {S.}~\bibnamefont
  {{Braginskii}}},\ }in\ \href@noop {} {\emph {\bibinfo {booktitle} {Reviews of Plasma Physics}}},\ \bibinfo {editor} {edited by\ \bibinfo
  {editor} {\bibfnamefont {M.A.}~\bibnamefont {Leontovitch, Vol.1}}}\ (\bibinfo  {publisher} {Consultants Bureau},\ \bibinfo {address} { New York},\ \bibinfo {year}
  {1965})\ pp.\ \bibinfo {pages} {205--311}\BibitemShut {NoStop}%
\bibitem [{\citenamefont {Alboussi\`ere}\ \emph {et~al.}(2019)\citenamefont
  {Alboussi\`ere}, \citenamefont {Drif},\ and\ \citenamefont
  {Plunian}}]{Alboussiere2019}%
  \BibitemOpen
  \bibfield  {author} {\bibinfo {author} {\bibfnamefont {T.}~\bibnamefont
  {Alboussi\`ere}}, \bibinfo {author} {\bibfnamefont {K.}~\bibnamefont {Drif}},
  \ and\ \bibinfo {author} {\bibfnamefont {F.}~\bibnamefont {Plunian}},\
  }{\bibfield  {journal}
  {\bibinfo  {journal} {Phys. Rev. E}\ }\textbf {\bibinfo {volume}
  {101}},\ \bibinfo {pages} {033107 } (\bibinfo {year} {2020})}\BibitemShut
  {NoStop}%
\bibitem [{\citenamefont {Priede}\ and\ \citenamefont
  {Avalos-Z{{\'u}}{{\~n}}iga}(2013)}]{Priede2013}%
  \BibitemOpen
  \bibfield  {author} {\bibinfo {author} {\bibfnamefont {J.}~\bibnamefont
  {Priede}}\ and\ \bibinfo {author} {\bibfnamefont {R.}~\bibnamefont
  {Avalos-Z{{\'u}}{{\~n}}iga}},\ }\href {\doibase
  https://doi.org/10.1016/j.physleta.2013.06.007} {\bibfield  {journal}
  {\bibinfo  {journal} {Physics Letters A}\ }\textbf {\bibinfo {volume}
  {377}},\ \bibinfo {pages} {2093 } (\bibinfo {year} {2013})}\BibitemShut
  {NoStop}%
\end{thebibliography}
%

\end{document}